## Catalyzing Informed Residential Energy Retrofit Decisions via Domain-Specific LLM

Lei Shu [a,b], Dong Zhao [a,b,c*], Jianli Chen [d], Armin Yeganeh [a], Sinem Mollaoglu [a], and Jiayu Zhou [e]

a. School of Planning, Design, and Construction, Michigan State University, East Lansing, Michigan, USA.
b. Human-Building Systems Lab, Michigan State University, East Lansing, Michigan, USA.
c. Department of Civil and Environmental Engineering, Michigan State University, East Lansing, Michigan, USA.
d. School of Civil Engineering, Tongji University, China
e. School of Information, University of Michigan, Ann Arbor, MI, USA

* Corresponding author: Dr. Dong Zhao, Professor and Graduate Director, Michigan State University, 552 W Circle Dr, East Lansing, Michigan, USA. Email: dz@msu.edu

### Abstract

Residential energy retrofit initiation is often stalled by an expertise gap, where homeowners lack the technical literacy required for structured building energy assessments and are thereby trapped in low-information environments with fragmented sources. To bridge this gap, this study reports a domain-specific large language model (LLM) designed to catalyze informed decision-making based solely on homeowner-accessible, natural-language descriptions, e.g., building age, size, and location. The model is created using the parameter-efficient low-rank adaption (LoRA) fine-tuning approach on a massive corpus grounded in physics-based energy simulations and techno-economic calculations from 536,416 U.S. residential building prototypes. Nine major retrofit categories are evaluated, including envelope upgrades, HVAC systems, and renewable energy installations. Validations against physics-grounded benchmarks show that the LLM consistently identifies high-quality retrofit options, achieving top-3 hit rates of 98.9% for maximum $CO_2$ reduction and 93.3% for the shortest discounted payback year. Moreover, the model exhibits strong robustness under incomplete input conditions, maintaining stable performance even when basic dwelling descriptions are only 60% partially specified. By significantly lowering the information activation energy for non-expert users while maintaining the scientific rigor, this physics-based AI model offers a scalable pathway for parallelized, user-centered decision making, accelerating cumulative energy savings and emission reductions across community and national scales.

### 1. Introduction

Residential energy retrofits serve as a primary lever for mitigating carbon emissions and advocating the global green building movement [1]. In the United States, the residential sector is responsible for approximately one-fifth of total energy consumption [2], positioning housing as a critical frontier of decarbonization. However, despite the environmental and economic benefits, retrofit adoption is hindered by systemic inertia stemming from a complex nexus of knowledge, behavioral, and financial barriers. Unlike commercial buildings that are managed by professional facility teams, residential retrofit decisions are primarily initiated by homeowners who lack specialized expertise in building energy assessment. Homeowners must navigate the non-trivial interactions among physical dwelling characteristics [3, 4], heterogeneous occupant behaviors [5, 6], and localized climate conditions [7, 8]. Furthermore, because residents spend nearly 90% of their time indoors [9], they often undergo thermal habituation, adapting to prevailing conditions rather than perceiving a deficiency that necessitates intervention. Coupled with high upfront capital requirements and the fact that energy expenditures represent 8–14% of household income [10], these factors create a high activation energy for decision-making, which discourages trial-and-error approach common in other consumer sectors.

A fundamental paradox in current retrofit landscape is that, although building experts and engineers possess tools to produce accurate building performance and economic estimates [11], homeowners rarely access this expertise due to the time and budgetary costs of on-site assessment. Consequently, residential retrofit decision-making remains trapped in a low-information environment, relying on fragmented sources such as anecdotal peer experiences, predefined policy incentives, or market-mediated professional advice [12, 13]. Information from peers is often dwelling-specific and fails to generalize across heterogeneous home conditions [8]. Similarly, policy incentives frequently prioritize administratively convenient measures over technically optimal solutions for specific dwellings [14]. Even professional consultations are often constrained by commercial incentives, particularly for small-scale projects with low profit margins [15, 16] where bias recommendations toward standard contractor offerings may take over dwelling-specific optimality. Therefore, the resulting disconnect between advanced analytic capabilities and the informal decision-making environment creates a critical need for a catalyst, i.e., a mechanism that can bridge the expertise gap and translate latent retrofit intent to informed action.

Large language models (LLMs) offer a transformative pathway to bridge this expertise gap by enabling a natural-language interface that allows homeowners to describe their dwellings and receive decision-relevant guidance [17, 18]. However, general-purposed LLMs commonly used in practice are trained on broad, non-domain-specific corpora and lack grounding in building physics and retrofit economics. This absence of engineering-level reasoning limits their reliability, often resulting in recommendations that lack numerical rigor or technical consistency [4].

Beyond model reliability, a secondary challenge lies in the cognitive architecture of the decision-making itself. In practice, homeowners rarely select measures based solely on technical optimality; instead, they must balance financial constraints, idiosyncratic preferences, and perceived risks. Therefore, decision-making is more appropriately framed as selecting from a small set of high-quality candidate options rather than a single prescribed solution. Offering multiple well-performing alternatives, supported by transparent performance metrics, can empower homeowners to evaluate technoeconomic trade-offs aligned with their household constraints.

To address these challenges, this study develops domain-specific LLM fine-tuned on a massive corpus of physics-based energy simulations and retrofit economic data. This LLM functions as a catalyst which transforms complex analytic outputs into accessible, actionable knowledge. The model inputs require only minimal, non-technical descriptions, such as building age, size, and location, to generate dwelling-specific retrofit recommendations. By design, the model identifies a small suite of high-quality retrofit candidates, each accompanied by estimated performance outcomes including $CO_2$ reduction, net site energy savings, total retrofit cost, and the discounted payback year (DPY). This approach ensures that the resulting decision support is both technically robust and behavioral flexible. Through a scalable, user-centered interface, this LLM significantly enhances homeowner decision-making capacity. It facilitates the transition from passive intent to active commitment, offering a viable mechanism to accelerate the adoption of effective retrofit measures and decarbonization promotion from the community to national scales.

## 2. Background

### *2.1. Data-Driven Models for Building Energy Retrofit Assessment*

Data-driven models have been increasingly used for building energy retrofit assessment, encompassing both surrogate models trained on physics-based energy simulation outputs and models developed using empirical building energy data [19].

Artificial neural networks (ANNs) have become a prominent class of surrogate models for estimating building energy performance and screening retrofit options. These models are typically trained on large sets of synthetic data generated from physics-based simulations, enabling them to approximate retrofit impacts with far lower computational cost than full EnergyPlus or TRNSYS runs. Several studies illustrate this surrogate-modeling paradigm. Ascione et al. trained ANNs on 1,350 EnergyPlus-generated cases to predict retrofit outcomes for Italian building categories [20]. Asadi et al. defined 950 Latin hypercube–sampled scenarios for a calibrated Portuguese residence, simulated each with TRNSYS, and trained an ANN to emulate performance [21]. Zhan et al. used a full year of EnergyPlus simulations—over 166,000 hourly records—for a Chinese campus building and developed an ANN that served as the basis for multi-objective retrofit optimization via NSGA-II [22]. At the urban scale, Thrampoulidis et al. simulated 12,806 Zurich dwellings with UBEM and trained ANN surrogates that map simple building descriptors to envelope retrofit solutions, reducing per-building computation to near-instantaneous levels [23]. Zhang et al. similarly trained ANNs using 10,368 HOT2000 simulations for a Canadian reference house to support residential retrofit decision analysis [24].

Beyond building energy performance emulation, ANN-based methods have evolved to address broader decision contexts including homeowner preferences, probabilistic adoption, and high-resolution time-series dynamics. Kaklauskas et al. developed an ANN-assisted decision system that uses catalog performance and cost data, weighted by user preferences, to rank millions of possible retrofit combinations for a Lithuanian passive house without requiring additional simulations [25]. Nyawa et al. trained an ANN classifier on the 51,000-household TREMI dataset to estimate the likelihood of homeowners undertaking retrofits, enabling targeted outreach and incentive allocation in France [26]. More recent studies integrate recurrent neural networks and attention mechanisms to incorporate measured temporal data. Nutkiewicz et al. combined EnergyPlus baselines with Long Short-Term Memory (LSTM) models trained on multi-year utility data to predict hourly energy use and evaluate retrofit effects for 29 U.S. mixed-use and commercial buildings [27]. Deb et al. applied an attention-based LSTM to sensor and utility data from a Swiss dwelling to forecast heating demand and identify cost-optimal retrofit packages [28].

Tree-based ensemble models such as Random Forest (RF), XGBoost, and LightGBM have become another dominant class of retrofit surrogates, offering strong predictive accuracy and scalability for both individual buildings and national building stocks. Araújo et al. used the 60,000 Energy Performance Certificates (EPCs) to train Extra Trees that predict EPC indicators and drive a budget-constrained optimizer for individual Portuguese dwellings [29]. Shan et al. varied eight envelope variables to generate 1,000 EnergyPlus simulations for a Chinese residential building, and trained LightGBM as the surrogate to support multi-objective envelope optimization [30]. At the national scale, Ali et al. combined four Irish residential archetypes with Latin hypercube sampling and automated EnergyPlus simulations to generate a synthetic stock of approximately one million dwellings, which was then used to train stacked XGBoost and LightGBM models for predicting end-use energy consumption and EPC labels at scale [31]. Several studies in China trained tree-based ensemble surrogates on a few hundred EnergyPlus simulations for representative school and residential buildings, using them to screen retrofit options and analyze trade-offs in energy, carbon, comfort, and cost [32-34]. Piras et al. combined recent measured consumption records with regional cost data in Italian context to train a RF model for annual energy demand and coupled it with economic indicators to identify the best retrofit measures [35]. Xu et al. trained causal-forest models on measured utility, weather, and retrofit data for 552 U.S. federal buildings to estimate heterogeneous savings for different retrofit action groups and guide portfolio-level prioritization [36]. Markarian et al. used EnergyPlus simulations on an archetypal Canadian office building to train ANN and tree-based surrogate models and coupled the best predictors with NSGA-II to derive Pareto-optimal retrofit packages that reduce peak load, energy use, and discomfort with substantially lower computational cost [37].

Despite their computational efficiency and strong predictive accuracy, ANN- and tree-based methods implicitly assume structured, expert-level inputs and specialized modeling workflows. These assumptions limit their applicability for homeowner-facing decision support, as non-expert homeowners are typically unable to provide the required technical inputs and directly interact with these tools. Consequently, such data-driven models are predominantly applied by researchers or energy professionals rather than serving homeowners.

### *2.2. LLMs to Support Building Energy Retrofit Decision*

LLMs have more recently been explored as tools to automate modeling workflows and provide more accessible, conversational interfaces for building energy applications and retrofits.

Perspective and review papers by Liu et al. [18] and Zhang et al. [38] outline how LLMs can support tasks such as automated code and model generation, document understanding, intelligent control, compliance checking and lifecycle data management, while highlighting challenges related to computational cost, data quality, hallucinations and domain adaptation. Within this research landscape, Jiang et al. proposed EPlus-LLM, which fine-tunes a Transformer model to translate natural-language building descriptions into EnergyPlus input files and automatically run simulations, thereby reducing manual modeling effort [39]. Xu et al. developed an LLM-based platform that connects building sensors, simulations, and a conversational agent to provide real-time feedback on health, comfort and energy use for occupants [40]. Choi and Yoon's GPT-UBEM framework uses GPT-4o to assist data preprocessing, feature engineering, energy prediction and scenario analysis for urban building stocks through natural-language prompts [41].

Recent studies have explored the use of LLMs to support residential retrofit decision-making and consistently indicate that LLMs are effective for early-stage, qualitative analysis but remain limited in providing reliable, quantitatively grounded decision support. Hidalgo-Betanzos et al. [42] evaluated general-purpose LLM chatbots for generating retrofit measures and identifying retrofit challenges, finding that these models can assist preliminary scoping but still require expert oversight to ensure technical correctness. Chen et al. [43] similarly assessed LLM-based retrofit assistance and showed that, although LLMs help structure retrofit considerations and identify potential challenges, their outputs are less reliable when used for ranking and decision-making beyond early-stage exploratory analysis. Shu et al. [4] further demonstrated that LLMs exhibit clear limitations from a techno-economic perspective, particularly in cost realism, payback estimation, and trade-off reasoning. A shared limitation across these studies is that the evaluated LLMs are general-purpose models whose outputs lack the numerical rigor and consistency required for engineering-level retrofit decision-making. As a result, while these models can support homeowner-facing discussion and early-stage reasoning, they lack quantitatively reliable and calibrated outputs required for actionable residential retrofit decisions.

In summary, widely explored data-driven models, such as ANNs and tree-based models, trained on physics-based energy simulation outputs or empirical building energy data can produce quantitatively reliable outputs, but their reliance on structured, expert-level inputs and technical workflows makes them unsuitable for homeowner-facing decision support. In contrast, general-purpose LLMs provide a more accessible, natural-language interface for homeowners, yet their quantitative outputs are not professionally calibrated against building physics and retrofit economics, limiting their reliability for engineering-level decisions. This study addresses this gap by fine-tuning an LLM on physics-based building energy simulation outputs and retrofit economic data, enabling homeowner-facing retrofit recommendations that are both reliable and easy to use.

## 3. Methodology

As illustrated in Figure 1, the methodology for our domain-specific LLM consists of three stages: dataset preparation, model fine-tuning, and model evaluation. In the first stage, physics-based energy simulations and techno-economic calculations generate diverse optimal retrofit outcomes across a large sample of buildings, while building parameter selection identifies homeowner-accessible building descriptors; these outputs are combined to construct the fine-tuning corpus, which is used for model training and evaluation. In the second stage, a base LLM is fine-tuned via Low-Rank Adaptation (LoRA) adapters to associate building descriptions with retrofit measures and their corresponding performance outcomes. In the final stage, the fine-tuned model is evaluated on retrofit selection tasks, focusing on its ability to identify optimal solutions within a small set of candidate measures. This evaluation reflects the intended use of the model as an early-stage decision support tool, where identifying high-quality candidate options is more critical than precise numerical prediction. The predicted performance outcomes (e.g., $CO_2$ reduction and DPY) are generated alongside the recommended measures and are used to support interpretation and comparison across options.

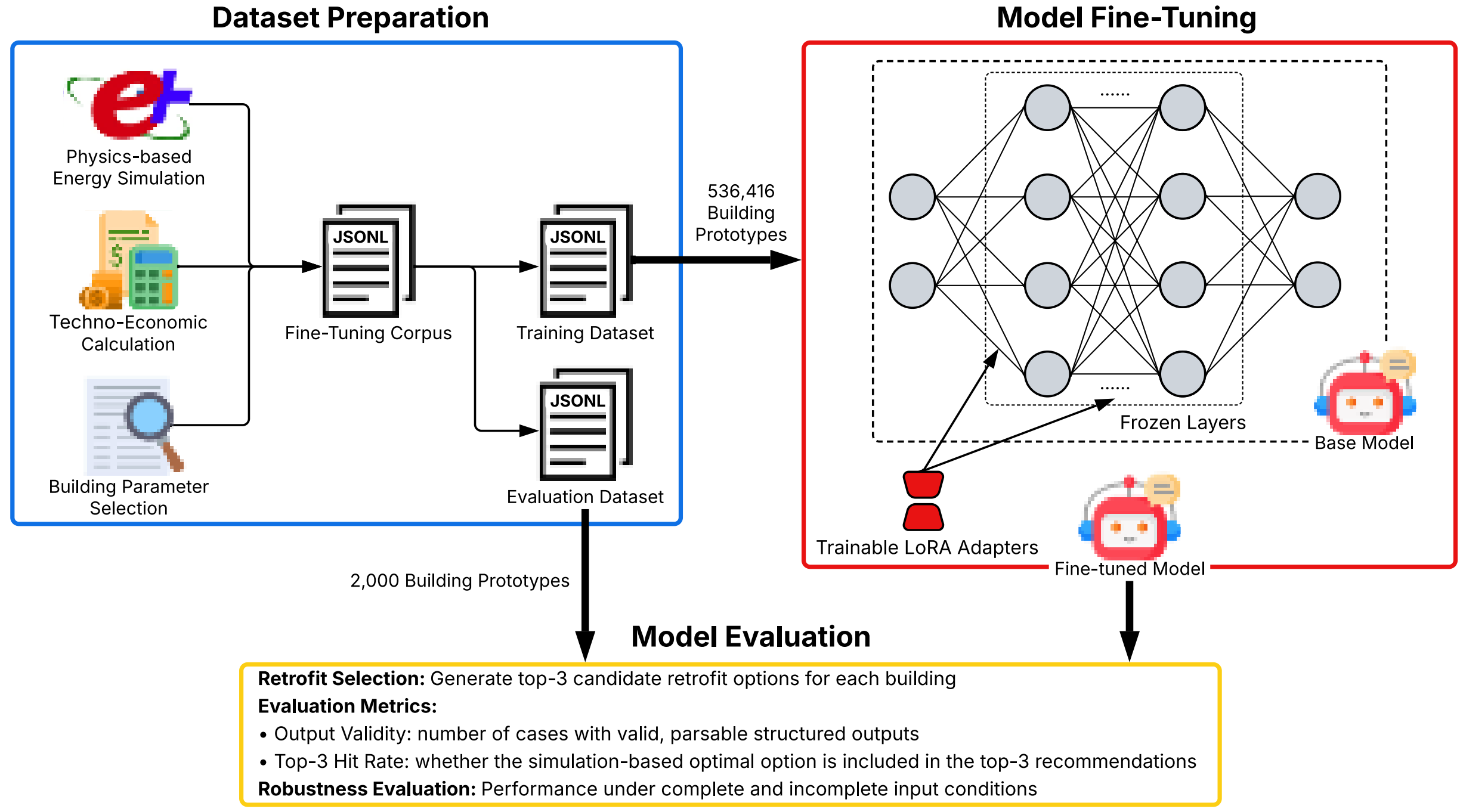

Figure 1. Workflow for domain-specific LLM development.

### *3.1. Dataset Preparation*

This section presents the dataset preparation used to support LLM fine-tuning (Figure 2). Section 3.1.1 introduces the data sources for physics-based energy simulation, economic calculation, and building parameter selection. Section 3.1.2 describes the data processing and transformation into the inputs and outputs of the fine-tuning dataset. Section 3.1.3 explains the organization of the dataset into a fine-tuning corpus for model training and evaluation. Section 3.1.4 summarizes the key characteristics and distributions of the dataset.

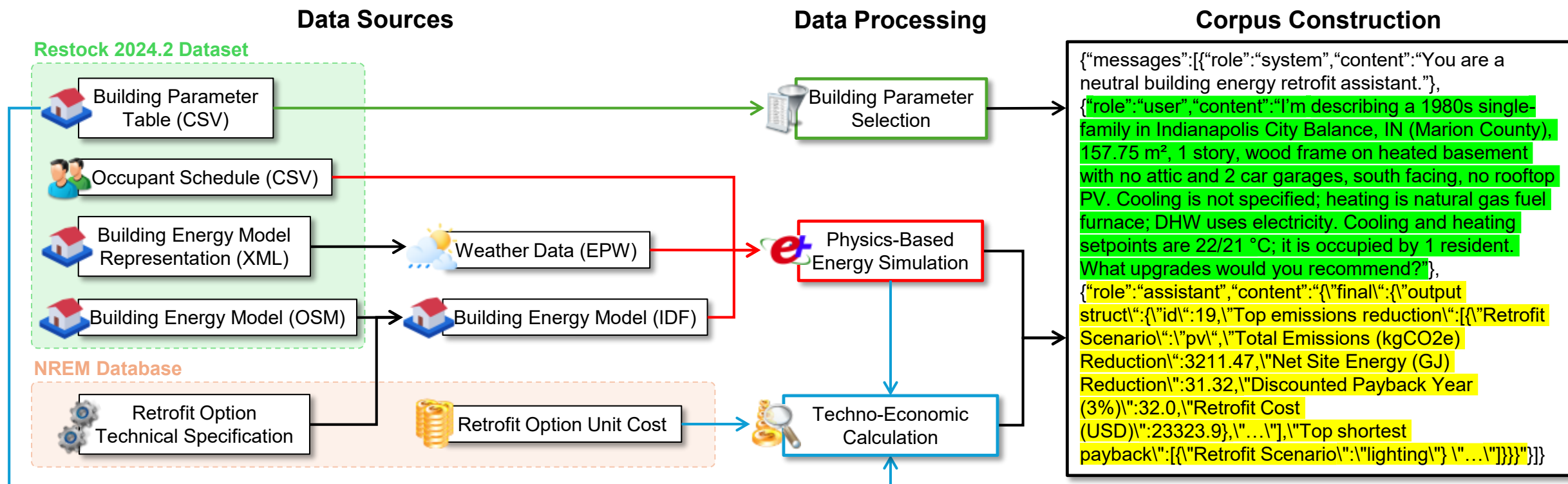

Figure 2. Dataset preparation workflow.

#### *3.1.1. Data Sources*

This study relies on two nationally representative data sources. Residential building prototype data were obtained from the ResStock 2024.2 dataset, developed by the National Laboratory of the Rockies (NLR) [44]. Retrofit measure data were drawn from the NLR’s National Residential Efficiency Measures (NREM) database [45].

Each building prototype is specified by building energy model files and associated parameter tables. The core building energy models are provided as OpenStudio Model (OSM) files and were batch-converted to EnergyPlus Input Data Files (IDFs) using Python-based workflows built on the OpenStudio SDK and the Eppy library. For each converted IDF file, an EnergyPlus Weather (EPW) file was assigned based on the weather identifier specified in the XML metadata associated with the corresponding OSM file [46]. Occupant schedules were supplied in tabular (CSV) format. Together, the IDF files, assigned EPW files, and occupant schedule data constitute the complete set of inputs for the physics-based energy simulations. The building prototype data also include parameter information stored in CSV format for techno-economic calculations and building parameter selection. These parameters describe location, building physical characteristics, system configurations, occupant-related usage patterns, utility rates, and $CO_2$ emission factors.

Retrofit measure data consist of technical specifications and unit cost data. Nine retrofit categories widely applicable to residential buildings were considered, including wall insulation, roof and ceiling insulation, window replacement, air sealing, HVAC upgrade, photovoltaic (PV) installation, appliance upgrade, lighting replacement, and water-heater upgrade. For each category, a representative retrofit option was selected to reflect an upper-bound performance level based on its primary technical specification (e.g., thermal resistance for insulation and efficiency coefficients for HVAC systems). The technical specifications of the selected retrofit options were used to modify the corresponding parameters in the building prototype IDF files, resulting in post-retrofit building energy models. In parallel, the associated unit cost data were combined with building physical characteristics and system configuration parameters from the building prototype data to support retrofit cost calculation. Table 1 summarizes the technical specifications and cost formulations for the nine representative retrofit options considered in this study.

*3.1.2. Data Processing*

For each building prototype, a baseline building energy model and multiple post-retrofit building energy models were simulated using EnergyPlus (version 24.2.0), which is a physics-based building energy simulation engine widely used in building energy analysis and retrofit evaluation studies [11]. The simulation timestep was set to one hour, corresponding to an hourly simulation resolution. This configuration struck a balance between capturing building energy performance with sufficient fidelity and avoiding the substantial computational burden of minute-scale simulations. Simulation outputs included the annual net site energy consumption for each building, along with the annual total consumption of electricity, natural gas, propane, and fuel oil reported separately by fuel type. All simulation results were exported as CSV files and subsequently post-processed in Python.

For each building prototype $i$, annual $CO_2$ emissions and annual energy costs were computed from simulated annual net site energy consumption by fuel type $C_{j,i}$, using the corresponding emission factors $EF_j$ and utility rates $UR_j$, as defined in Eqs. (1) and (2), where $j$ indexes fuel types. Baseline and post-retrofit results were compared to obtain $CO_2$ emission reductions, net site energy reductions, and energy cost savings for each retrofit option.

$$E_{CO2,i} = \sum_{j} C_{j,i} \times EF_j \tag{1}$$

$$Cost_i = \sum_{j} C_{j,i} \times UR_j \tag{2}$$

The DPY was calculated by comparing the retrofit cost $I$ with cumulative discounted annual energy cost savings $S_t$ over post-retrofit years $t$, as defined in Eq. (3). A constant annual discount rate $d$ of 3% was applied to future energy cost savings. The retrofit cost $I$ was calculated according to the cost formulations summarized in Table 1.

$$DPY = \min\left\{ n \;\middle|\; \sum_{t=1}^{n} \frac{S_t}{(1+d)^t} \geq I \right\} \tag{3}$$

Based on the performance outcomes derived from the physics-based energy simulations and techno-economic calculations, retrofit options were ranked separately for each building prototype under different criteria. Specifically, the top-3 retrofit options achieving the largest $CO_2$ emission reductions and the top-3 retrofit options with the shortest DPY were identified. For each identified retrofit option, the associated performance outcomes, including $CO_2$ emission reduction, net site energy reduction, energy cost saving, and DPY, were recorded.

To construct the fine-tuning dataset, the computed retrofit outcomes described above were combined with simplified building descriptions derived from the building prototype tabular data. For each building prototype, a subset of building parameters was selected to form the input representation, consisting of 21 parameters that summarize building characteristics typically known to homeowners, such as building age, type, floor area, number of stories, primary HVAC system type, heating fuel, and thermostat setpoints. These building descriptions were paired with the corresponding ranked retrofit options and their associated performance outcomes.

Table 1. Baseline and retrofit specifications, parameter modifications, and cost calculation methods for the nine retrofit options.

| No. | Retrofit option | Modified parameter(s) | Retrofit value(s) | Unit | Cost calculation |
|---|---|---|---|---|---|
| 1 | Wall insulation | Wall material thermal conductivity | Thermal conductivity = Thickness / 6.34 | W/m·K | Wall insulation cost = 150.4 * exterior wall area |
| 2 | Roof & ceiling insulation | Roof/ceiling material thermal conductivity | Thermal conductivity = Thickness / 8.63 | W/m·K | Roof/ceiling insulation cost = 19.7 * roof area |
| 3 | Window replacement | Glazing U-value; SHGC | U-value = 1.476; SHGC = 0.22 | $W/m^2 \cdot K$ | Window replacement cost = 974.4 * total window area |
| 4 | Air sealing | Effective leakage area (ELA) | ELA = 0.08 * baseline ELA | $m^2$ | Air sealing cost = 11.8 * conditioned floor area |
| 5 | HVAC upgrade – DX cooling + DX heating (electric ASHP/MSHP) | DX cooling COP; DX heating COP | Cooling COP = 8.32; Heating COP = 3.2 | – | Number of units = round(cooling capacity / 7.03); HVAC cost = number of units * 1623 + 634.41 |
| | HVAC upgrade – DX cooling only | DX cooling COP | Cooling COP = 7.39 | – | Number of units = round(cooling capacity / 7.03); HVAC cost = number of units * 1623 + 634.41 |
| | HVAC upgrade – Electric furnace / baseboard | Heating efficiency (COP-equivalent) | Heating COP = 3.2 | – | Number of units = round(heating capacity / 3.52); HVAC cost = number of units * 1699 + 634.41 |
| | HVAC upgrade – Natural gas furnace | Burner efficiency | Burner efficiency = 0.98 | – | Heating units = round(heating capacity / 3.52); Cooling units = round(cooling capacity / 25.11); HVAC cost = heating units * 1699 + cooling units * 3472.93 + 2217.75 |
| | HVAC upgrade – Fuel furnace (oil / propane / other) | Burner efficiency | Burner efficiency = 0.80 | – | Heating units = round(heating capacity / 3.52); Cooling units = round(cooling capacity / 39.32); HVAC cost = heating units * 1699 + cooling units * 3232 + 2217.75 |
| | HVAC upgrade – Hot-water boiler (shared heating) | Boiler thermal efficiency | Boiler efficiency = 0.95 | – | Heating units = round(heating capacity / 3.52); Cooling units = round(cooling capacity / 21.98); HVAC cost = heating units * 1699 + cooling units * 3472.93 + 4399.77 |
| | HVAC upgrade – Shared cooling system | DX cooling COP | Cooling COP = 8.32 | – | Number of units = round(cooling capacity / 7.03); HVAC cost = number of units * 3073 + 2309.70 |
| 6 | Photovoltaic (PV) installation | Cell efficiency; active area fraction; inverter efficiency | Cell efficiency = 0.21; Active area fraction = 0.22; Inverter efficiency = 0.95 | – | PV capacity = roof area * 0.22 * 0.21 * 1000. PV unit cost: if capacity < 880 W, unit cost = 4.30; if between 880 and 14080 W, unit cost = (4.37 − 0.000091 * PV capacity); if > 14080 W, unit cost = 3.10. PV cost = PV capacity * unit cost |
| 7 | Appliance replacement | Appliance power scaling | Refrigerator * 0.76; Washer * 0.333; Dishwasher * 0.76; Dryer * 0.9467 (electricity) or 0.9456 (gas) | – | Appliance cost = 1159.02 (refrigerator) + 1350.76 (washer) + 1079.79 (dishwasher) + 453.69 (dryer), applied only if the appliance is replaced |
| 8 | Lighting replacement | Interior lighting power | Lighting power * 0.47 | – | Number of fixtures = round(conditioned floor area / 6.97); Lighting cost = number of fixtures * 7.87 |
| 9 | Water-heater upgrade | Rated COP of heat pump water heater | 4.07 | – | Water heater replacement cost = 3707 (fixed value) |

Note:
SHGC = solar heat gain coefficient; ELA = effective leakage area; DX = direct expansion; COP = coefficient of performance; ASHP = air-source heat pump; MSHP = mini-split heat pump; PV = photovoltaic system.
Costs are in USD; areas in $m^2$; cooling and heating capacities in kW; PV capacity in W; PV unit cost in USD/W. Round() denotes round-half-up rounding, with a minimum of one unit for non-zero conditioned floor area.

*3.1.3. Corpus Construction*

A fine-tuning dataset comprising 538,416 residential building prototypes was constructed. From this dataset, 2,000 building records were randomly selected and held out as an evaluation set, while the remaining records were used for model fine-tuning. All building prototype records were transformed from structured tabular formats into a JSONL corpus following a System–User–Assistant schema, where the system message specifies a neutral building energy retrofit assistant role, the user message provides a homeowner-accessible building description identified through the building parameter selection process, and the assistant message reports the corresponding top-3 retrofit options and associated performance outcomes derived from physics-based energy simulations and techno-economic calculations under predefined optimization objectives.

To enhance linguistic diversity and reduce sensitivity to a single phrasing style, 14 natural-language templates were designed to express the descriptions using different sentence structures, information ordering, and phrasing styles. The templates preserve identical semantic content while varying only surface-level linguistic form. Each building record was assigned to one template according to predefined sampling weights, ensuring that multiple language styles are represented across the corpus while each building appears only once.

*3.1.4. Dataset Characteristics*

Figure 3 summarizes the distribution of building layout-related attributes in the building prototype dataset. Single-family detached buildings constitute the dominant building type. In terms of foundation, slab is the most prevalent configuration. For attic types, no attic is the most common condition, while vented attic also represents a substantial share. Regarding garage configuration, the majority of prototypes do not include a garage. Overall, the dataset covers multiple layout categories while remaining dominated by a few common configurations.

Figure 4 shows the distribution of climate zones in the building prototype dataset. Zone 5A accounts for the largest share, corresponding to a cold-humid climate (e.g., Chicago). Zone 4A follows, representing a mixed-humid climate (e.g., New York). Zone 3A represents the next largest share and corresponds to a warm-humid climate (e.g., Atlanta). Zone 2A follows, corresponding to a hot-humid climate (e.g., Houston). Zone 3B, representing a warm-dry climate (e.g., Los Angeles), also accounts for a notable share. The remaining zones account for relatively small shares. This distribution reflects the underlying residential building stock represented in the ResStock dataset. The dataset is constructed using stock-weighted sampling based on the U.S. housing inventory and is therefore not uniformly distributed across climate zones.

Figure 5 shows the distribution of floor area in the building prototype dataset. The 0–46 $m^2$ range accounts for the largest share, which is largely associated with individual units in multi-family buildings. As floor area increases, the proportion of building prototypes decreases steadily.

Figure 6 shows the distribution of building vintage in the building prototype dataset. Buildings from the 1970s account for the largest share, with relatively consistent proportions from the 1980s to the 2000s. A substantial portion of the dataset consists of buildings constructed before 1980, indicating strong representation of older building stock.

Together, these distributions provide a comprehensive characterization of the dataset and ensure a diverse and representative foundation for subsequent model training and evaluation.

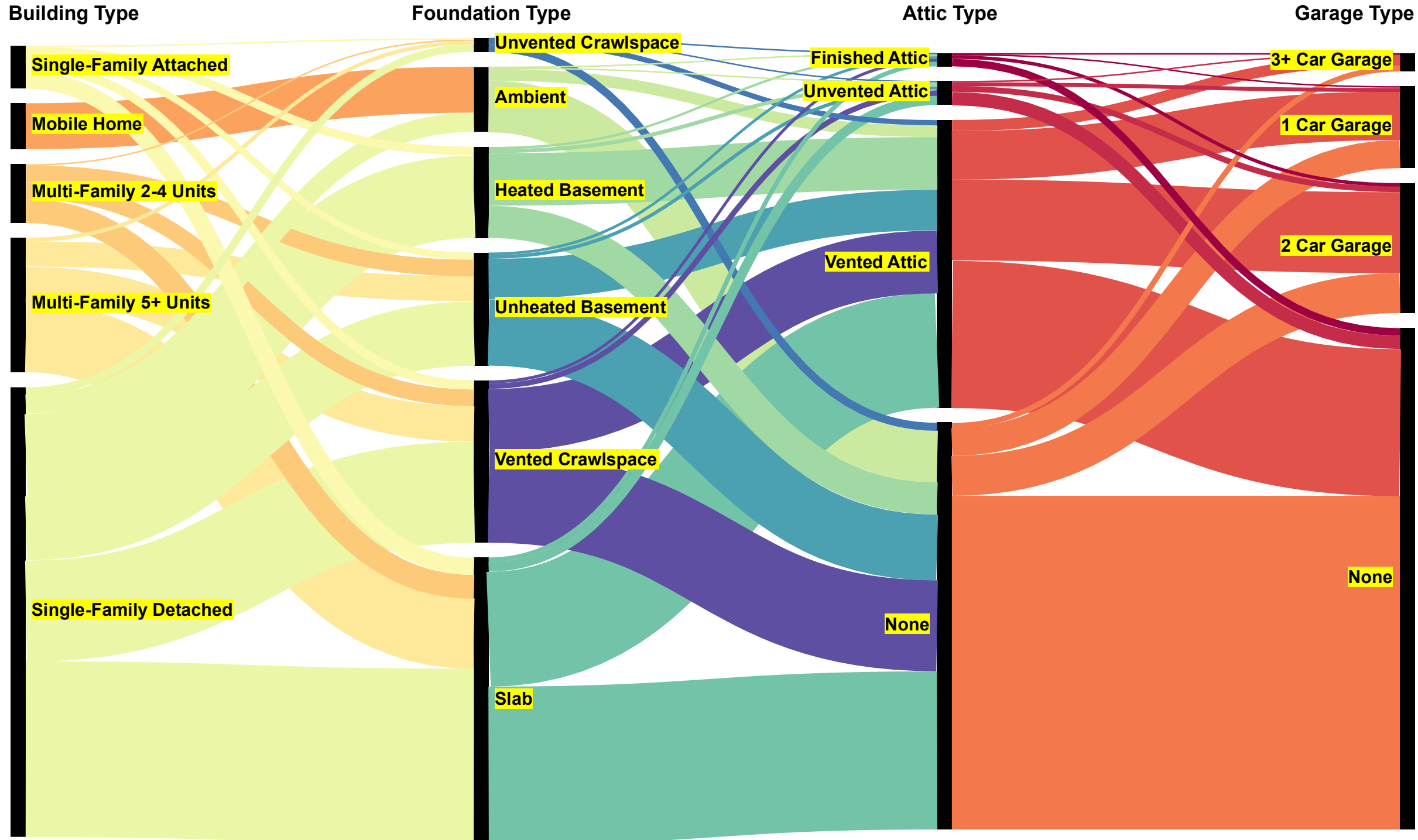

Figure 3. Distribution of building layout in the dataset.

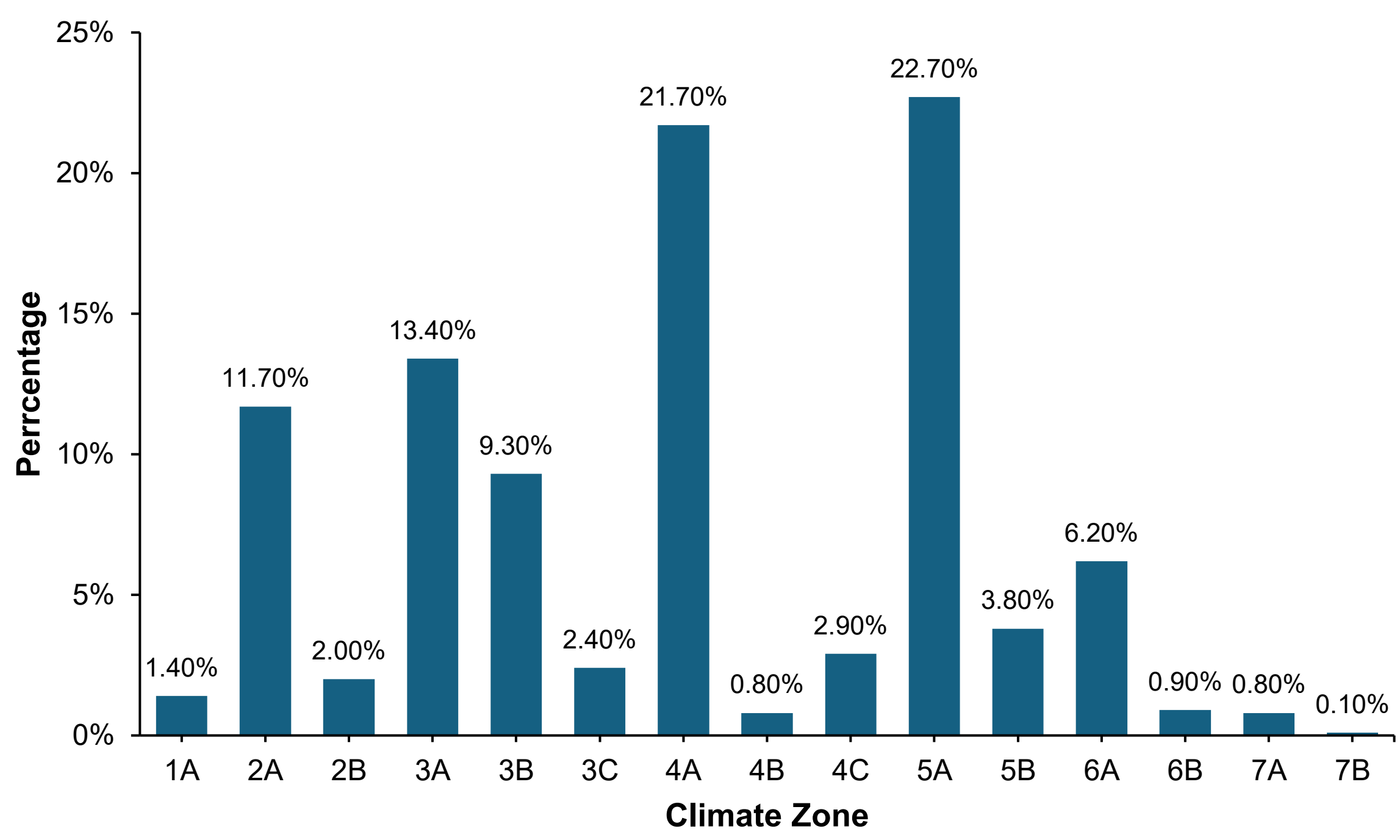

Figure 4. Distribution of climate zones in the dataset.

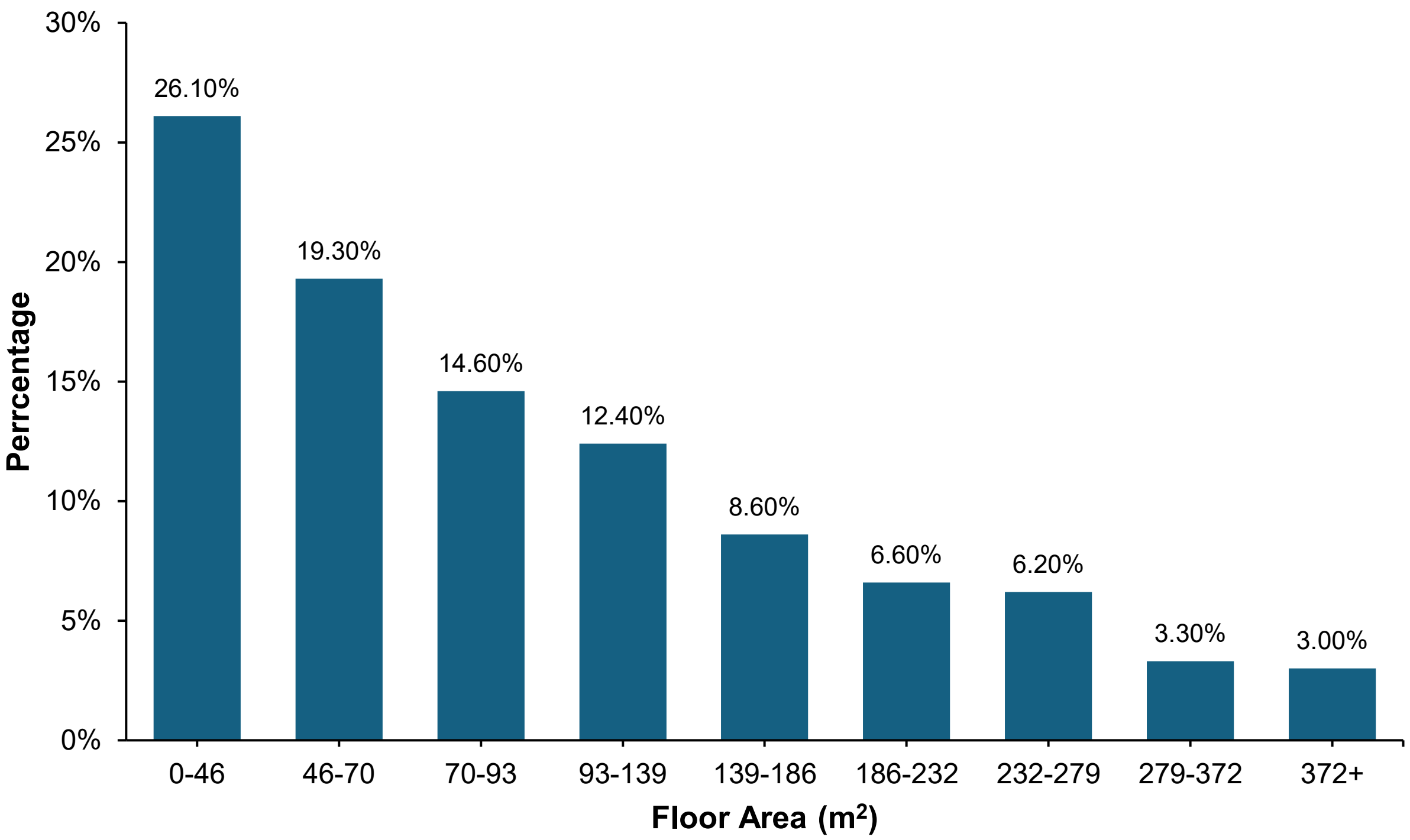

Figure 5. Distribution of floor area in the dataset.

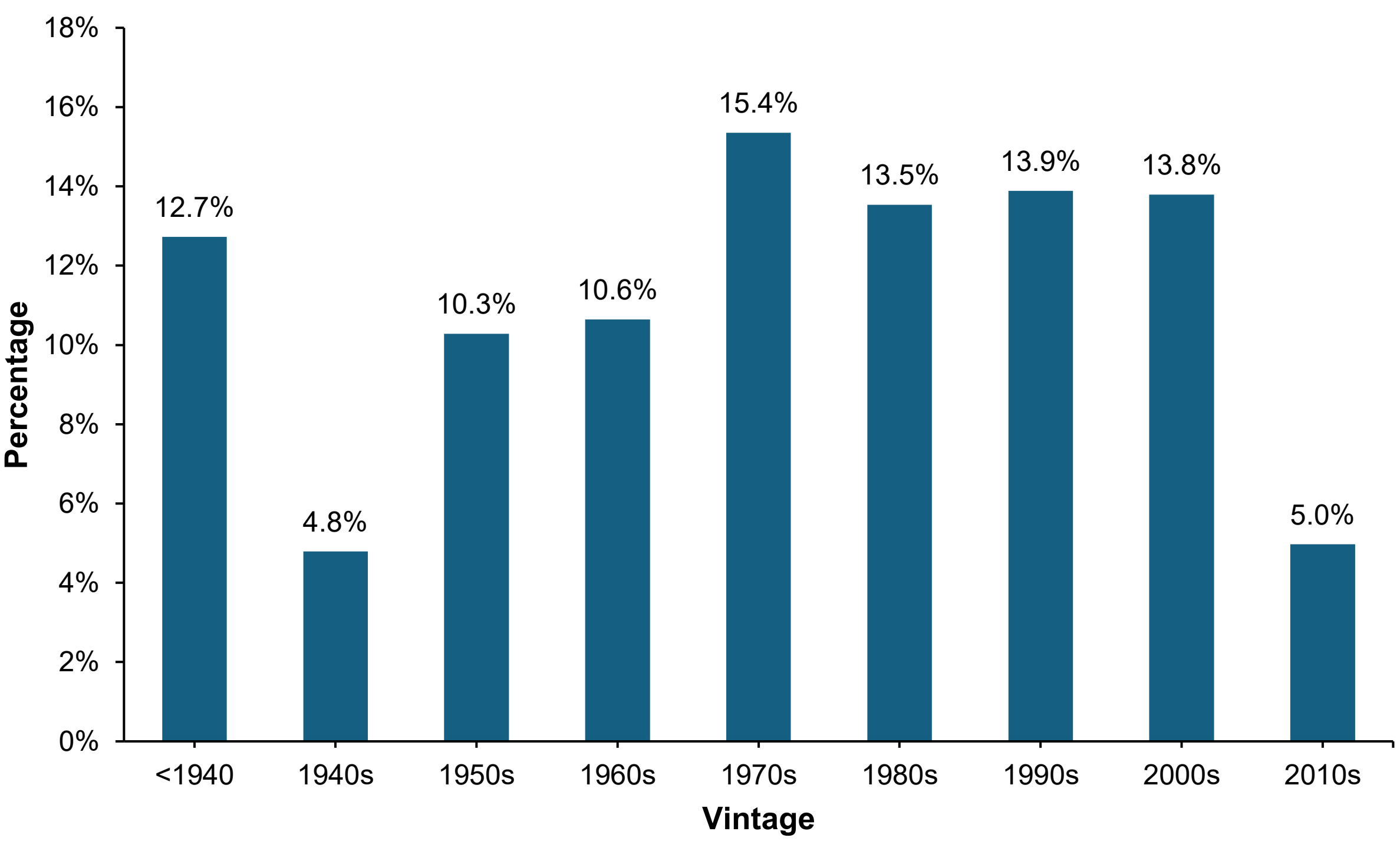

Figure 6. Distribution of building vintage in the dataset.

### *3.2. Model Fine-Tuning*

The base LLM was Qwen3-8B-Base, a transformer-based, decoder-only LLM with approximately 8 billion trainable parameters [47]. Fine-tuning was performed to adapt this base model to residential retrofit decision-making, by training the model to associate natural-language building descriptions with computation-derived optimal retrofit outcomes.

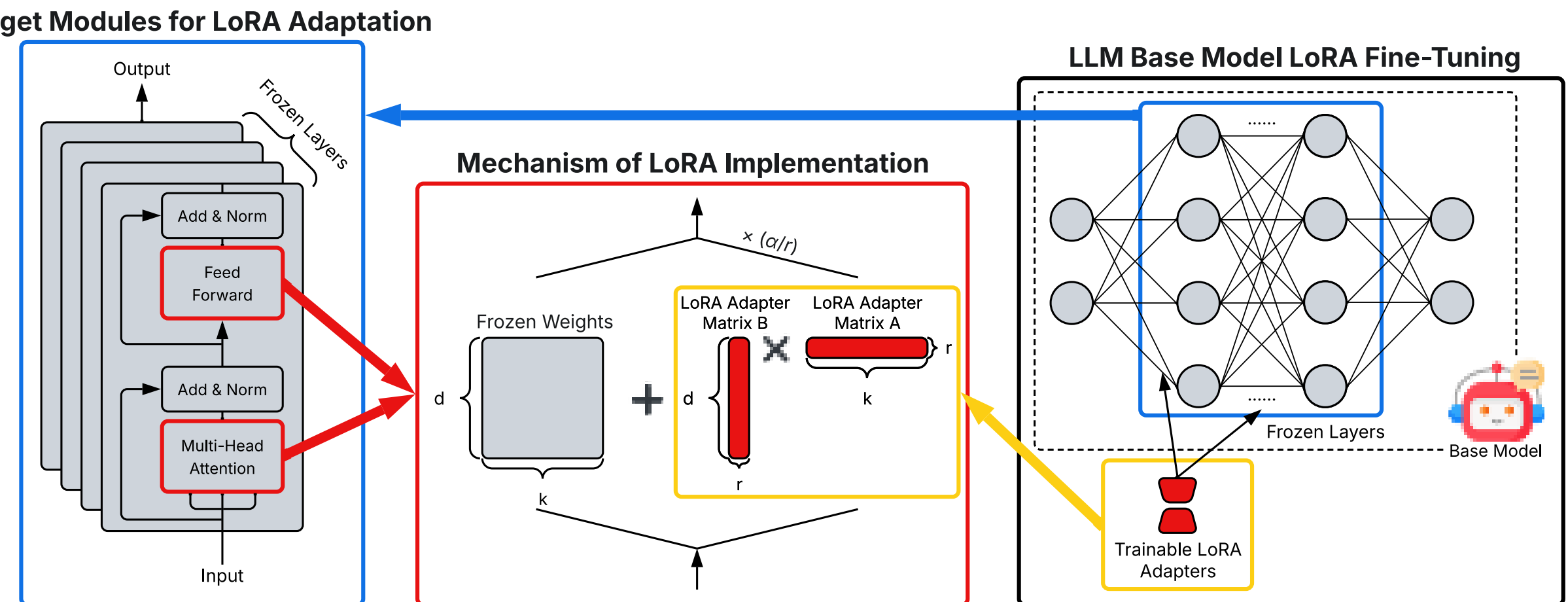

Figure 7. LoRA fine-tuning framework.

Fine-tuning was implemented using LoRA, a parameter-efficient approach that enables domain adaptation while preserving the original model parameters [48]. LoRA introduces trainable low-rank decomposition matrices into selected modules within the frozen layers of the base model. As illustrated in Figure 7, these adapters (Matrix A and Matrix B) introduce low-rank weight updates that are added to the frozen base model weights without modifying the original model parameters. In this study, LoRA adapters were applied to both the multi-head attention modules, which determine how the model weighs and relates different parts of the input description, and the feed-forward modules, which transform the attended information into higher-level representations used for output generation.

The LoRA rank $r$ was set to 16, representing a balance between adaptation capacity and training stability: lower values constrain the model's ability to capture domain-specific patterns, whereas higher values increase flexibility but also raise the risk of overfitting and unnecessary computational cost. The scaling factor $\alpha$ was set to 32 to modulate the influence of the LoRA weights on the model's internal representations. By applying the standard scaling factor ($\alpha/r$), the magnitude of the low-rank updates remains consistent, ensuring that the adaptation is neither too subtle to affect behavior nor too aggressive to destabilize training. Additionally, a dropout rate of 0.05 was applied to mitigate overfitting by encouraging distribution of learning across multiple trainable LoRA weights within each layer, rather than relying disproportionately on only a few.

Fine-tuning was performed at the Michigan State University's High Performance Computing Center using one NVIDIA A100 graphics processing unit (GPU) equipped with 64 GB memory. To improve computational efficiency and reduce GPU memory consumption for long input sequences, mixed-precision computation was employed [49]. Forward and backward computations were performed using bfloat16 precision to improve training efficiency, while FP32 precision was used during parameter updates for the trainable LoRA adapters to ensure numerical stability. The optimization objective followed a completion-only strategy: loss was calculated solely on the response portion of each training sample, ensuring that the model learns only the target retrofit recommendations.

### *3.3. Model Evaluation*

Model evaluation in this study focuses on the model's ability to support retrofit decision-making under practical usage conditions. Two aspects are considered: output validity and retrofit selection performance.

The evaluation workflow is summarized in Figure 8, which illustrates the prompt structure, the organization of model outputs, and the evaluation pipeline linking model generations to physics-grounded simulation results. During evaluation, model outputs were required to follow a strict and predefined JSON schema. This structured output requirement enabled reliable script-based extraction of model outputs and ensured automated and reproducible evaluation.

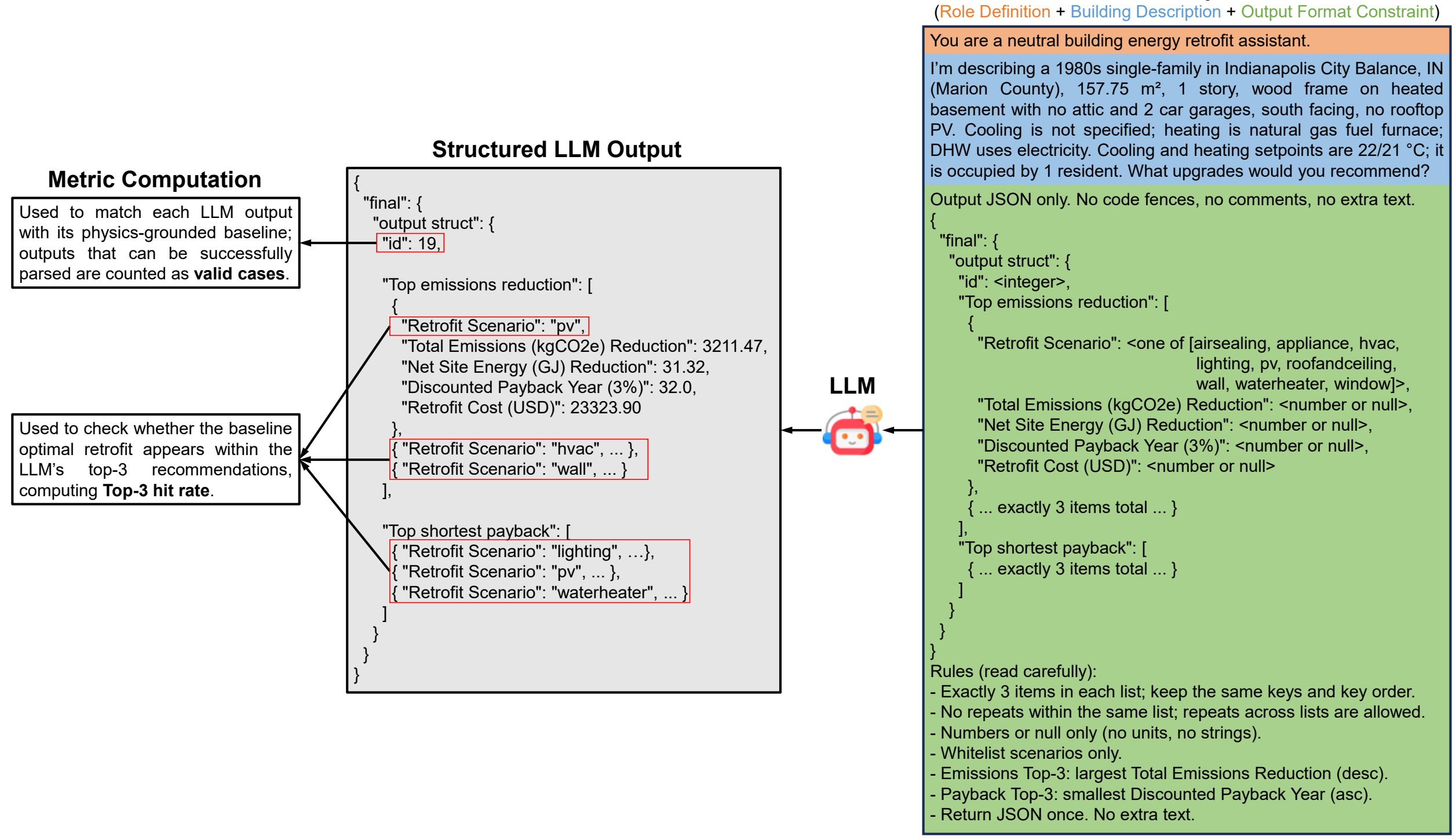

Figure 8. Model evaluation workflow.

Output validity was evaluated using the number of valid cases. A valid case is defined as a building for which the model output can be successfully parsed according to the predefined JSON schema. This metric reflects the model's ability to generate structured and usable outputs, which is essential for downstream evaluation.

Retrofit selection performance was evaluated using the top-3 hit rate, which measures whether the optimal retrofit option identified by physics-based simulation appears within the model's top-3 recommendations. The metric is calculated as the fraction of valid cases for which this condition is satisfied. This metric reflects the model's ability to generate a small set of high-quality candidate measures that contains the optimal solution. Compared to stricter single-choice metrics, Top-3 hit rate better aligns with the intended use of the system as an early-stage decision support tool, where identifying a reliable candidate set is more practically meaningful than selecting a single exact option.

Model robustness was examined under two input conditions. In the complete-input condition, the model receives the full set of homeowner-accessible building parameters used during fine-tuning. In the incomplete-input condition, 0%–40% of the non-core building parameters are randomly masked to simulate partial information availability, while core parameters such as building type, floor area, and building age are retained. The same evaluation metrics were applied under both conditions to assess the stability of model performance when input information is incomplete.

## 4. Results

### *4.1. Overall Distribution of Optimal Retrofit Measures*

Figure 9 compares the distribution of the top-3 optimal retrofit measures under two optimization objectives: maximizing $CO_2$ reduction and minimizing DPY. For $CO_2$ reduction, PV installation (27%), water-heater upgrade (22%), and HVAC upgrade (21%) together account for approximately 70% of the selections, showing a strong concentration on system-level measures. For DPY, the distribution shifts, with lighting replacement (36%) and PV installation (31%) together accounting for about 67%. Lighting replacement is characterized by low upfront cost, while PV installation involves higher investment but yields substantial energy savings.

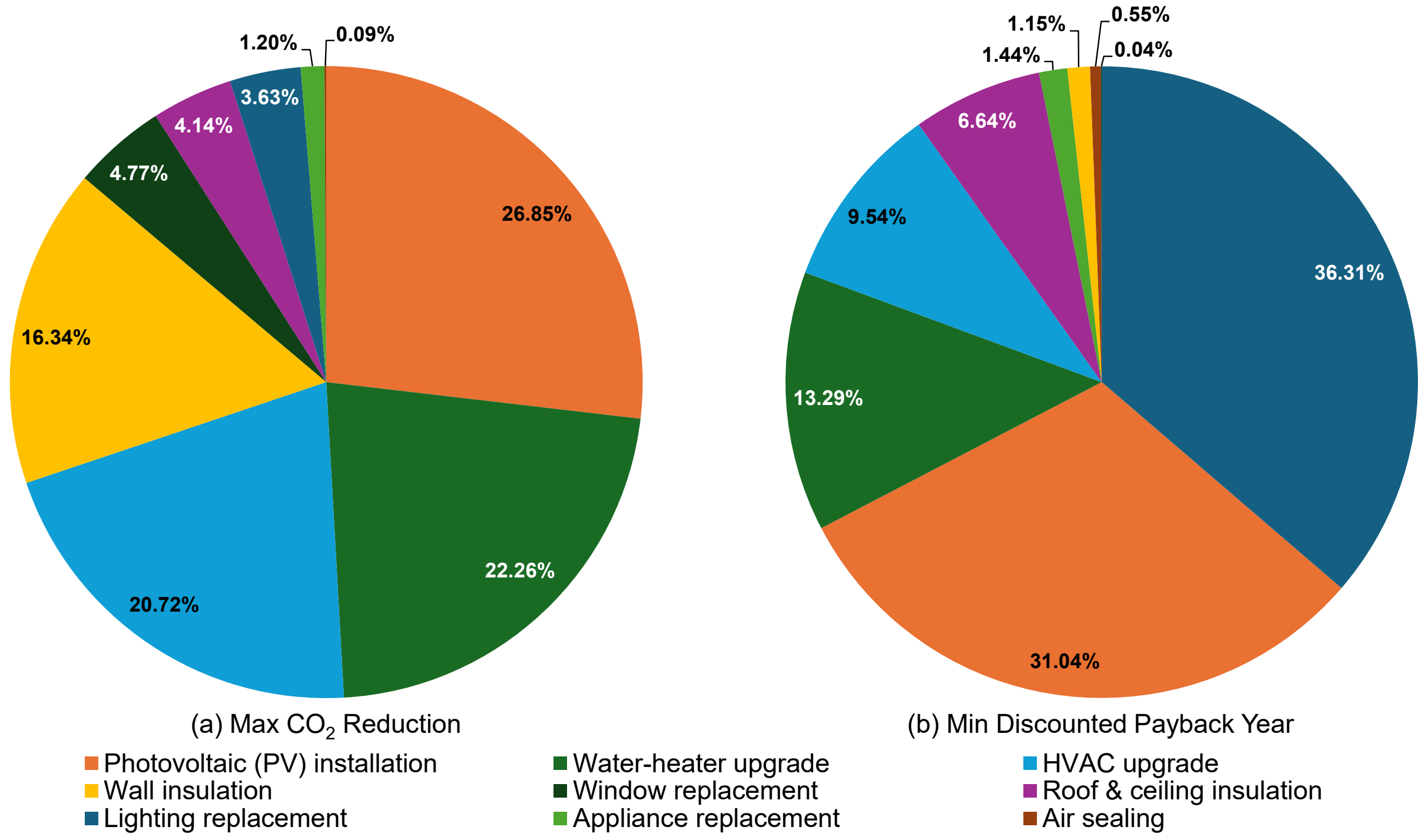

Figure 9. Distribution of top-3 optimal retrofit measures under $CO_2$ reduction and DPY objectives.

### *4.2. Homeowner-Facing Natural-Language Interface*

Figure 10 shows the graphical interface used to demonstrate how the developed LLM supports residential retrofit decision-making from a homeowner perspective. The interface illustrates the core interaction paradigm used in this study: users provide a free-text description of basic dwelling characteristics using natural language, and the LLM returns ranked retrofit measures together with associated performance outcomes.

In the "Enter House Description" panel, the input consists only of basic building information that homeowners typically know. This design directly reflects the study's objective of lowering the input barrier for advanced retrofit decision-making. In the "Recommended Measures" panel, the LLM outputs ranked retrofit measures under different optimization objectives, together with the associated performance outcomes.

In this study, evaluation focuses on the top-3 hit rate of optimal retrofit selection, reflecting whether the most effective solution is included within the recommended set. The predicted performance values (e.g., $CO_2$ reduction and DPY) are provided to support user interpretation and comparison across options. This design positions the system as an early-stage decision support tool, where users can interpret the results and make final decisions based on their own preferences and constraints.

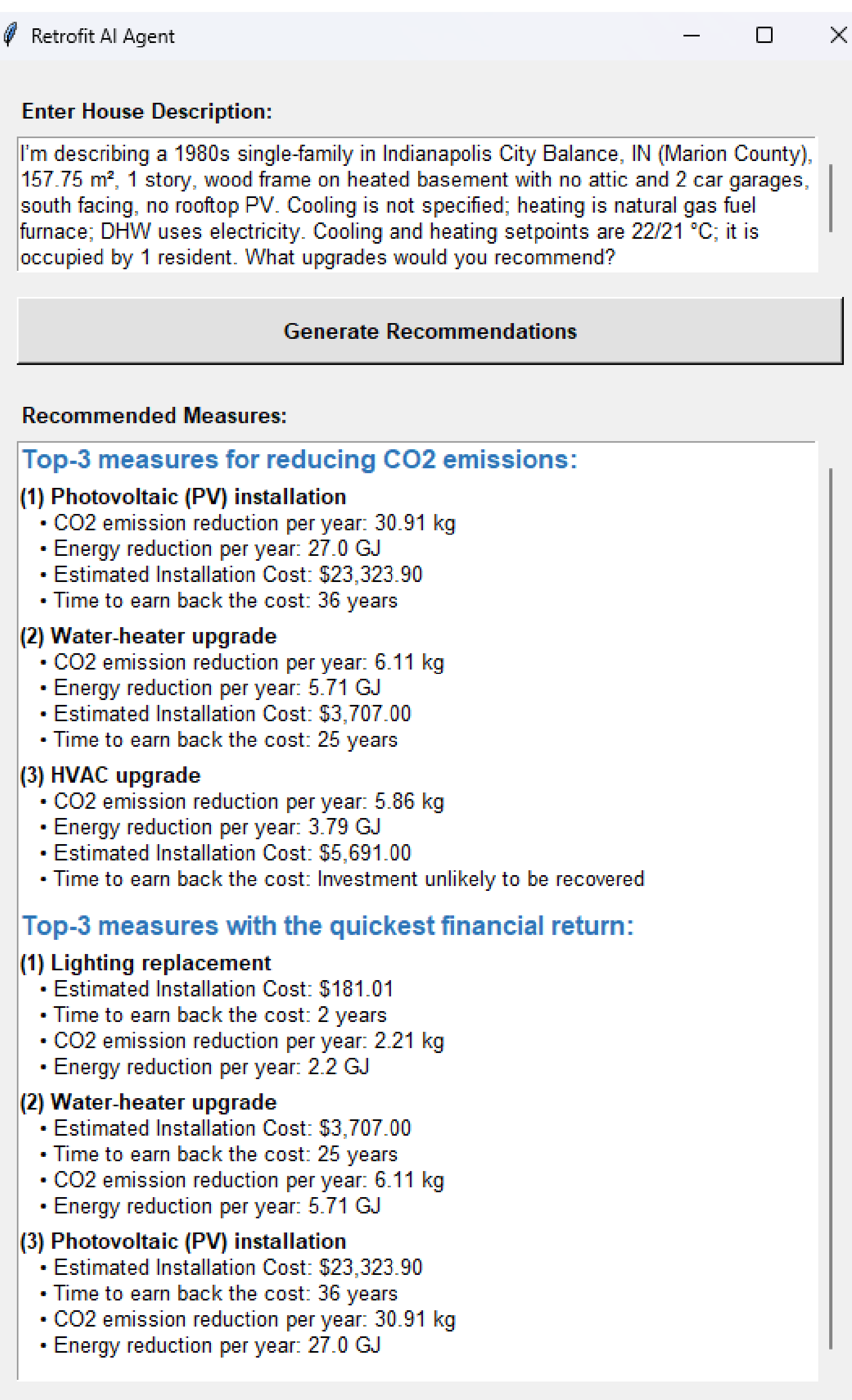

Figure 10. Screenshot of a homeowner-facing interface.

### *4.3. Valid Cases*

Figure 11 summarizes the number of valid cases used to evaluate retrofit selection under complete-input and incomplete-input conditions. As shown in this figure, the base LLM produces more outputs that can be directly parsed and evaluated, whereas the fine-tuned LLM produces fewer valid cases. The higher valid-case count for the base LLM reflects stricter adherence to the predefined output format, while the lower count for the fine-tuned LLM primarily arises from formatting deviations rather than incorrect retrofit selections.

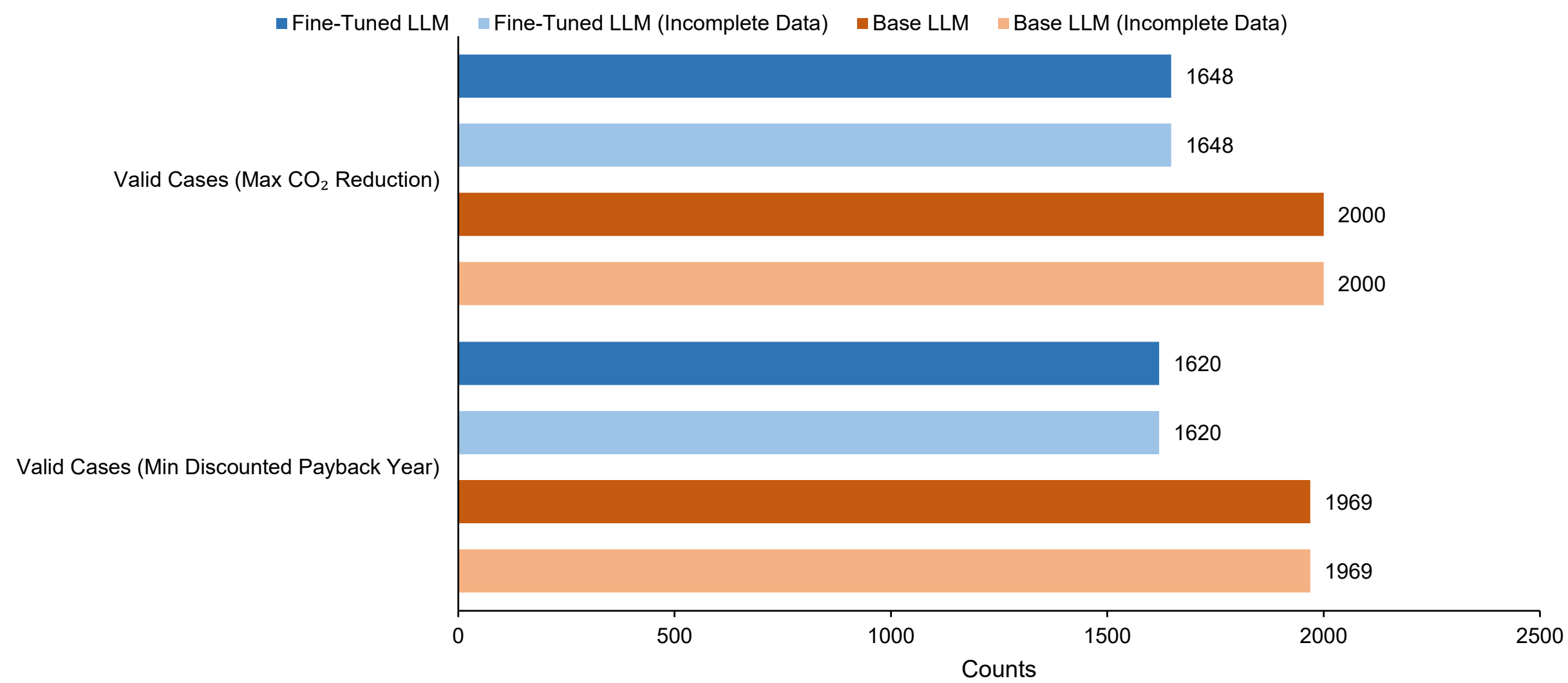

Figure 11. Number of valid cases used for model evaluation under complete-input and incomplete-input conditions.

Under incomplete-input conditions, the number of valid cases remains similar for both models, indicating that input incompleteness has a limited impact on output parsability compared to formatting deviations introduced during generation.

### *4.4. Optimal Retrofit Selection Performance*

Fine-tuning substantially and consistently improves optimal retrofit selection performance under both complete-input and incomplete-input conditions. Figures 12 (a) and (b) report the top-3 hit rate of the fine-tuned and base LLMs when selecting optimal retrofit measures under maximum $CO_2$ reduction and minimum DPY objectives.

For the $CO_2$ reduction objective, the fine-tuned LLM achieves consistently high top-3 hit rates, reaching 98.9% under complete input and 98.5% under incomplete input. In comparison, the base LLM reaches 79.3% under complete input and 82.4% under incomplete input. These results indicate that the fine-tuned LLM can reliably identify a small set of candidate retrofit measures that includes the optimal solution.

For the DPY objective, fine-tuning leads to substantial performance gains. The top-3 hit rate increases from 49.8% to 93.3% under complete input and from 64.8% to 93.2% under incomplete input. This improvement highlights the effectiveness of fine-tuning in capturing cost-related trade-offs, which are inherently more complex than energy or emission-based objectives.

Across both optimization objectives, the top-3 hit rate of the fine-tuned LLM exhibits only marginal degradation under incomplete-input conditions, indicating strong robustness to incomplete information. In contrast, the base LLM achieves slightly higher performance under incomplete-input conditions. This increase may be attributed to the reduced input complexity, which simplifies the decision space.

Importantly, the top-3 hit rate reflects the model's ability to provide a reliable set of high-quality candidate measures, rather than to identify a single exact solution. This aligns with the intended role of the system as an early-stage decision support tool, where users are presented with a small number of promising options and make final decisions based on their own preferences and constraints.

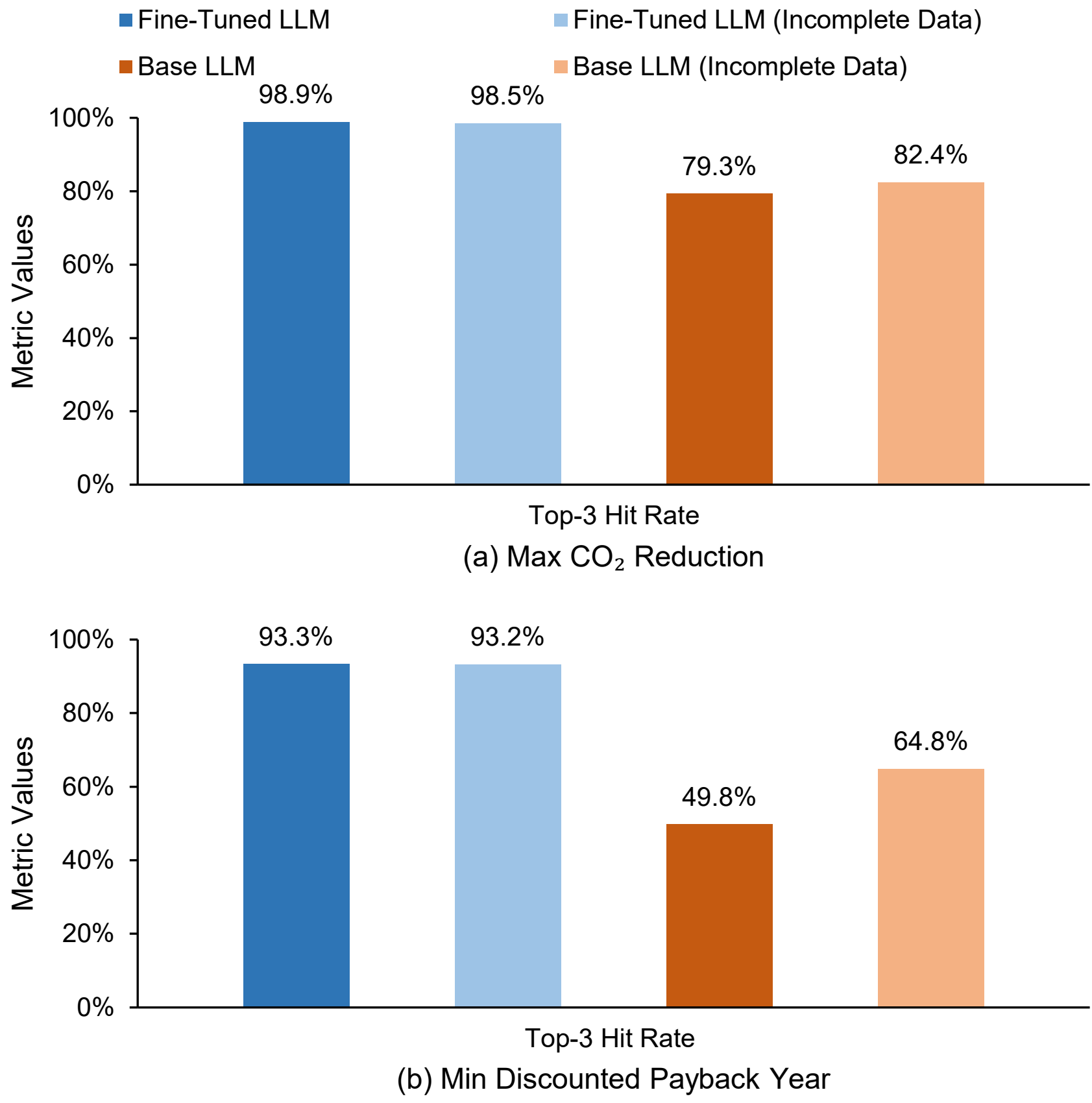

Figure 12. Top-3 hit rate of optimal retrofit measure selection under $CO_2$ reduction and DPY objectives with complete and incomplete inputs.

## 5. Discussion

### *5.1. Informed Decisions for Homeowners*

The primary contribution of this LLM lies in its role as a catalyst that lowers the threshold for engagement in residential retrofit. Conventional tools operate on the assumption that homeowners can navigate expert-driven modeling workflows, essentially requiring them to act as amateur building engineers. In contrast, our model allows retrofit planning to initiate directly from informal, natural-language descriptions. This shift eliminates the need for homeowners to first surmount the technical hurdle of structured data entry, allowing decision-relevant analysis to begin at the moment of intent formation rather than after a formal audit. Prior research shows that when early engagement in retrofit decision-making increases from approximately 0% to 10% of households in a community, the overall adoption of retrofit measures at the community level rises sharply, highlighting the catalytic effect of lowering the threshold for engagement [13].

This catalytic effect is most pronounced under conditions of incomplete information—the standard state of residential decision-making. By explicitly operating within these constraints and delivering calibrated, dwelling-specific recommendations from partial inputs, the model renders the decision process tractable at an earlier stage. Furthermore, our model addresses pervasive barrier of subjective uncertainty regarding upfront financial risk. Homeowners often hesitate not because potential benefits are unknown, but because they cannot assess whether those benefits meaningfully apply to their own dwelling. By presenting a small set of ranked retrofit options together with quantitatively grounded estimates of $CO_2$ reduction, energy savings, cost, and DPY, the model reduces subjective uncertainty while accommodating diverse homeowner preferences and constraints, supporting more evidence-based decisions. In this way, the proposed model functions as a decision-support mechanism that stabilizes retrofit choices under uncertainty.

### *5.2. Catalyzing Large-Scale Retrofit Adoption*

Our LLM facilities a shift from sequential knowledge diffusion to parallel activation across the community and national scales. Traditional retrofit adoption relies on a slow trickle-down of information from isolated pilot demonstrations or early local adopters. Our LLM bypasses this bottleneck by enabling thousands of households to simultaneously access accurate suggestions. This shift to distributed decision-making is critical as it effectively decentralizes expertise, moving it from the hands of a few professionals into the informal environments where residential choices are actually made. This fits well scalable retrofit adoption as large-scale retrofit performance increases when more households receive reasonable information for their informed decisions [13]. By lowering the barrier to accessing actionable, dwelling-specific knowledge, the model increases the number of homeowners who are able to independently evaluate and implement suitable retrofit measures. As a result, decision-making is no longer concentrated among a small subset of participants but is distributed across the broader population.

In addition, the model improves the alignment between individual decisions and system-level outcomes. By providing calibrated estimates of $CO_2$ reduction, energy savings, cost, and DPY, the model enables homeowners to select options that are both individually feasible and environmentally effective. This reduces the likelihood of misaligned retrofit choices that may arise from incomplete or generalized information.

As a result, the proposed model supports more effective aggregation of individual retrofit actions, leading to enhanced energy and emission outcomes across both community and national scales. From a policy perspective, this suggests that scalable, user-centered decision-support tools can complement or outperform traditional pilot-based programs in achieving large-scale retrofit impact.

### *5.3. Mechanisms and Pathways for Performance Improvement*

The catalytic efficiency of the model stems from scientific, rigorous fine-tuning on physics-based building energy simulations. Unlike general-purpose LLMs, which rely on implicit and often inconsistent associations learned from broad corpora, fine-tuning aligns the model's internal representations with domain-relevant techno-economic relationships. This alignment enables the model to consistently retrieve high-performing retrofit measures within a small candidate set, improving the reliability of decision support rather than merely enhancing linguistic plausibility.

In addition to parameter fine-tuning, the observed robustness under incomplete input highlights prompt engineering as an important pathway for practical performance improvement. Results indicate that modest reductions in input completeness lead to only minor degradation in top-3 candidate coverage, while maintaining a reliable set of high-performing retrofit options. This suggests that domain-specific prompts can be designed to require fewer or coarser building descriptions while still eliciting a stable pool of candidate retrofit measures. Such prompt strategies are particularly valuable in early-stage decision-making, where users benefit from exploring multiple viable retrofit options under limited information.

At inference time, retrieval-augmented generation (RAG) provides a complementary mechanism for improving domain-specific performance by incorporating external and updatable information. Through retrieval, the model can access updated retrofit measure specifications, regional emission factors, local energy prices, or user-provided building records, without repeated retraining [50]. When integrated with a domain-specific LLM, such retrieval mechanisms can help maintain quantitative relevance as policy incentives, grid emission intensities, and utility rates evolve, while preserving the calibrated decision logic learned during fine-tuning. This hybrid architecture highlights a promising direction in which fine-tuned domain knowledge and real-time contextual data are combined to support adaptive retrofit decision-making.

Finally, advances in multimodal LLMs indicate additional opportunities for extending domain-specific retrofit decision support. Visual inputs such as exterior building images or interior photographs may provide complementary cues about envelope conditions, window types, or system configurations that are difficult for homeowners to describe accurately in text alone. Recent work in related infrastructure domains demonstrates that images can be used to assess physical conditions [51], suggesting the potential for analogous applications in residential retrofit analysis. Incorporating visual information alongside natural-language descriptions may further reduce input burden and improve early-stage decision support, particularly when formal audits or detailed building records are unavailable.

### *5.4. Limitations and Future Work*

Several limitations of the present study should be acknowledged. First, the proposed domain-specific LLM was fine-tuned exclusively using U.S. residential building prototypes and retrofit measures. Its direct applicability to other national or regional contexts has not yet been established. Second, the retrofit decision space considered in this study is represented by nine representative retrofit options, each selected to reflect the upper-bound performance within its corresponding retrofit category. While this design choice enables consistent benchmarking and clear comparison across retrofit categories, it does not fully capture the diversity of options available in real-world markets. Future work may extend the proposed framework by incorporating region-specific data, expanding the range of retrofit alternatives, and integrating external information sources through mechanisms such as RAG. While RAG can enrich available options and contextual data, fine-tuning remains essential for learning the decision logic required to compare alternatives and identify high-performing candidate measures.

## 6. Conclusion

This study developed and validated a domain-specific LLM designed to act as a catalyst for residential energy retrofit decision-making. The LLM was fine-tuned on physics-based energy simulations, techno-economic calculations, and homeowner-accessible building descriptors derived from 536,416 residential building prototypes across the U.S. Requiring only natural-language descriptions of basic dwelling characteristics as inputs, our model outputs enable homeowners to bypass the technical friction of traditional energy modeling. The model successfully bridges the expertise gap that frequently stalls homeowner actions when facing decisions for optimal retrofit measures from nine major categories: wall insulation, roof and ceiling insulation, window replacement, air sealing, HVAC upgrade, PV installation, appliance replacement, lighting replacement, and water-heater replacement. For each recommended retrofit, the model also provides associated performance outcomes, including $CO_2$ reduction, net site energy reduction, retrofit cost, and DPY.

Evaluation results demonstrate that the model consistently identifies high-quality retrofit options within a small candidate set. The optimal solution for $CO_2$ reduction is included in the top-3 recommendations in 98.9% of cases, while the shortest-DPY solution is included in 93.3% of cases. In addition, the model exhibits strong robustness under incomplete input conditions, maintaining stable recommendation performance even when building information is partially specified.

The contribution of this study is twofold. First, by providing a small set of high-quality retrofit options together with transparent performance information, it enables homeowners to make immediate and flexible decisions. Users can evaluate multiple well-performing alternatives based on their own preferences and constraints, thereby reducing decision friction and increasing confidence in decision-making. Second, by reducing input requirements through natural-language interaction, the model makes decision support accessible to non-expert users, allowing a large number of homeowners to initiate decisions in parallel. Together, these contributions provide a scalable pathway for accelerating homeowner adoption of retrofit measures and achieving cumulative energy savings and emission reductions at community and national levels.

## CRediT authorship contribution statement

**Lei Shu:** Writing – review & editing, Writing – original draft, Visualization, Software, Methodology, Investigation, Formal analysis, Data curation, Conceptualization. **Dong Zhao:** Writing – review & editing, Supervision, Resources, Funding acquisition, Conceptualization. **Jianli Chen:** Writing – review & editing, Validation. **Armin Yeganeh:** Writing – review & editing, Validation. **Sinem Mollaoglu:** Writing – review & editing, Validation. **Jiayu Zhou:** Writing – review & editing, Validation.

## Declaration of competing interests

The authors declare that they have no known competing financial interests or personal relationships that could have appeared to influence the work reported in this paper.

## Acknowledgement

This study was supported by the National Science Foundation (NSF) of the United States through Grant #2046374. Any opinions, findings, conclusions, or recommendations expressed in this material are those of the researchers and do not necessarily reflect the views of NSF.

## Data availability

Data will be made available on request.